\documentclass[aps,noshowpacs,amsmath,amssymb,twocolumn]{revtex4}
\usepackage{amsmath,amssymb}
\usepackage{graphics,epsfig}
\usepackage{graphicx}
\usepackage{color}

\def\build#1_#2^#3{\mathrel{\mathop{\kern 0pt#1}\limits_{#2}^{#3}}}

 \newcommand{\vct}[1]{{\mbox {\boldmath $#1$}}}

\begin{document}

\title{Lagrangian time correlations of vorticity alignments in isotropic turbulence: observations and model predictions}

\author{Laurent Chevillard$^1$}
\author{Charles Meneveau$^2$}
\affiliation{$^1$Laboratoire de Physique de l'\'Ecole Normale Sup\'erieure
de Lyon, CNRS, Universit\'e de Lyon, 46 all\'ee d'Italie F-69007 Lyon,
France}
\affiliation{$^2$Department of Mechanical Engineering and Center
for Environmental and Applied Fluid Mechanics, The Johns Hopkins
University, 3400 N. Charles Street, Baltimore, MD 21218,
USA}

\begin{abstract}
Motivated by results from recent particle tracking experiments in turbulence (Xu et al., Nat. Phys. \textbf{7}, 709 (2011)), we study the Lagrangian time correlations of vorticity alignments with the three eigenvectors of the deformation-rate tensor. We use data from direct numerical simulations (DNS), and explore the predictions of a Lagrangian model for the velocity gradient tensor. We find that the initial increase of correlation of vorticity direction with the most extensive eigen-direction  observed  by Xu et al. is reproduced accurately using the Lagrangian model, as well as the evolution of correlation with the other two eigendirections. 
Conversely, time correlations of vorticity direction with the eigen-frame of the pressure Hessian tensor show differences with the model. 
\end{abstract}

 
\maketitle
In a recent communication \cite{XuPum11}, the inertial range  vorticity vector defined as the rotational motion of an ensemble of four Lagrangian particles in a turbulent flow (``tetrad" \cite{ChePum99}) has been shown, at short times, to display growing alignment and correlation with the eigenvector associated to the most positive eigenvalue of the associated initial deformation tensor. It is well established \cite{Ash87,Cantwell92,Men11} that at any given time the alignment between vorticity is most likely to be with the intermediate strain-rate eigenvector. However, how said correlation changes as function of time-delay between vorticity and strain-rate is an aspect of turbulence fine-scale structure that had not received much attention before. The time evolution of alignment correlations provides an interesting observable on which  turbulence models can be tested. The evolution of alignments between material lines and vorticity,  and viscous mechanisms of  tilting of vorticity, have been studied in experiments and simulations \cite{Gualaetal05,Holzneretal10}. 

In this Letter, we examine the Lagrangian time correlation of vorticity and the various eigenvectors of the deformation rate tensor, as predicted by a Lagrangian stochastic model \cite{CheMen06,Men11}. Results are compared with the results of  \cite{XuPum11} and additional data from Direct Numerical Simulation (DNS) of the Navier-Stokes equations in moderate Reynolds number isotropic turbulence.  In the same spirit as the analysis proposed in Ref. \cite{XuPum11}, we also study the short time alignment of vorticity with the pressure Hessian eigen-vector directions. The pressure Hessian tensor is a key quantity in the transport equation of the velocity gradients $\textbf{A}$, where $A_{ij} = \partial u_i/\partial x_j$ and  $\textbf{u}$ is the velocity vector, given by 
\begin{equation}\label{eq:TranspA}
\frac{d\textbf{A}}{dt} = -\textbf{A}^2-\textbf{P}+\nu\Delta \textbf{A}\mbox{ ,}
\end{equation}
where $d/dt$ is the Lagrangian time derivative, $P_{ij} = \partial p/\partial x_i\partial x_j$ the pressure Hessian (divided by fluid density) and $\nu$ the fluid's kinematic viscosity. The results obtained from DNS are compared with the predictions of the Lagrangian stochastic model of Ref. \cite{CheMen06}. 
  
As in Ref. \cite{XuPum11}, we focus now on the Lagrangian time correlation of vorticity $\vct{\omega} = \vct{\nabla}\wedge \textbf{u}$ with eigen-vectors of the rate of deformation tensor $\textbf{S} = (\textbf{A} +\textbf{A}^\top)/2$. We define, as in Ref. \cite{XuPum11}, the unit-norm vorticity vector direction $\textbf{e}_\omega(t) = \vct{\omega}(t)/|\vct{\omega}(t)| $ and the orthonormal eigenframe $(\textbf{e}_1(t),\textbf{e}_2(t),\textbf{e}_3(t))$ of $\textbf{S}$ associated to its 3 ordered eigenvalues $\lambda_1(t) >\lambda_2(t)>\lambda_3(t)$. Incompressible fluids are considered, so that $\lambda_1(t) +\lambda_2(t)+\lambda_3(t)=0$ which implies that $\lambda_1(t) \ge 0$ and  $\lambda_3(t) \le 0$. Following Ref. \cite{XuPum11}, who focused on the most extensive eigen-direction ($\textbf{e}_i(t)$ with $i=1$), the following Lagrangian correlation function is considered
\begin{equation}\label{eq:Cit}
C_i(\tau) = \langle [\textbf{e}_i(t).\textbf{e}_{\omega}(t+\tau)]^2\rangle\mbox{ ,}
\end{equation}
for $i\in\{1,2,3\}$ and the temporal displacement $\tau$ (and averaging in time $t$) is performed along Lagrangian trajectories.

\begin{figure*}[t]
\center{\epsfig{file=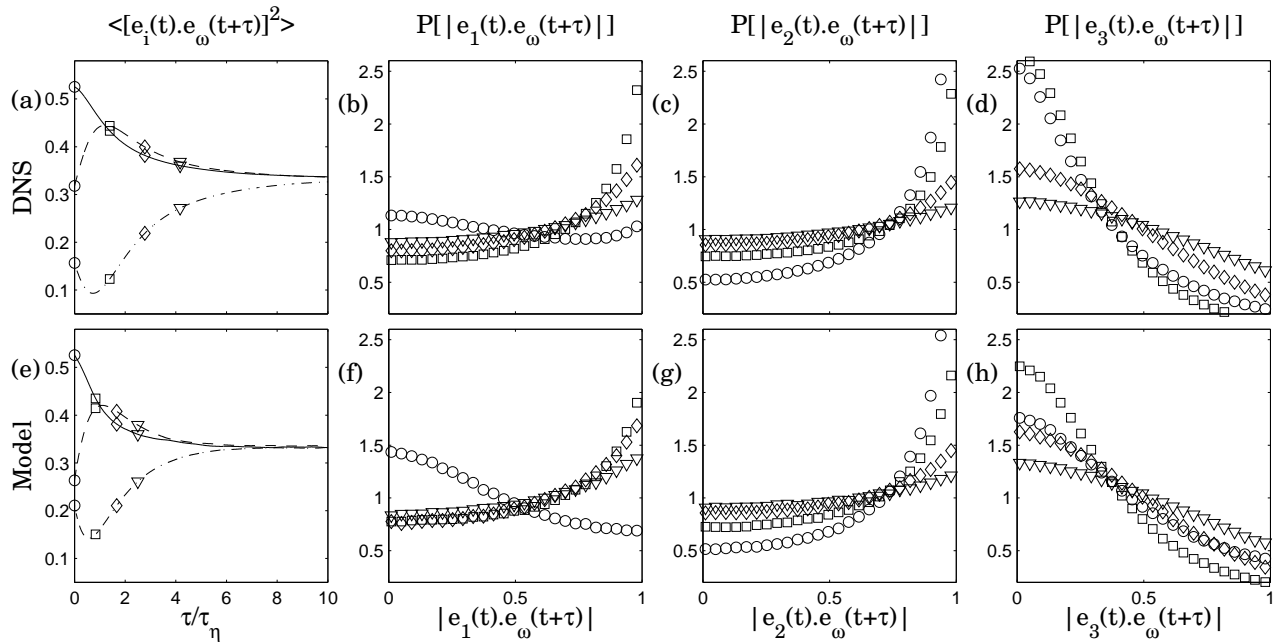,width=17cm}} \caption{(a) and (e) Correlation $C_i(\tau)$ (Eq. \ref{eq:Cit}) of the vorticity direction $\textbf{e}_{\omega}(t)$ and the eigenframe of the deformation rate, $\textbf{e}_i(t)$ in DNS flows and from the model (Eq. \ref{eq:Determourmodel}): dashed line for maximal straining direction (i.e. $i=1$), solid line for intermediate straining direction ($i=2$), and dot-dashed line for most contractive eigendirection ($i=3$). (b,c,d) and (f,g,h) Probability Density Functions (PDFs) of $|\textbf{e}_i(t).\textbf{e}_{\omega}(t+\tau)|$ at various time lag $\tau$ represented by different symbols, i.e. $0=\tau_{\circ}<\tau_{\square}<\tau_{\diamond}<\tau_{\triangledown}$ and indicated in (a) and (b). } \label{fig:CorrelAlignVortStr}
\end{figure*}

We consider predictions from the  Lagrangian model developed in Ref. \cite{CheMen06} for the velocity gradient tensor in 
turbulent flows and  given explicitly by the following stochastic differential equation
\begin{equation}\label{eq:Determourmodel}
d\textbf{A} =  \left(-\textbf{A}^2+
\frac{\mbox{Tr}(\textbf{A}^2)}{\mbox{Tr}(\textbf{C}_{{\tau_k}}^{-1})}
\textbf{C}_{{\tau_k}}^{-1}
-\frac{\mbox{Tr}(\textbf{C}_{{\tau_k}}^{-1})}{3T}
\textbf{A}\right)dt+d\textbf{W}\mbox{ .}
\end{equation}
The second and third terms in the right hand side of Eq. \ref{eq:Determourmodel} are closures for, respectively, (minus) the pressure Hessian and viscous Laplacian that govern the time evolution of $\textbf{A}$, as it can be seen by comparing with Eq. \ref{eq:TranspA} (see Ref. \cite{Men11} for a review of this and other models). The term $\textbf{W}$ is a tensorial delta-correlated noise term that has been added in order to represent possible forcing 
effects, e.g. from neighboring eddies. The ``recent Cauchy-Green tensor" $\textbf{C}_{\tau_k}$, which arises after invoking 
the ``recent fluid deformation'' approximation \cite{CheMen06,CheMen08},  can be expressed in terms of  matrix exponentials:
\begin{equation}\label{eq:MyCG}
\textbf{C}_{\tau_k} = e^{\tau_k \textbf{A}}e^{\tau_k
\textbf{A}^\top}\mbox{ ,}
\end{equation}
where $\tau_k$ is the Kolmogorov time-scale  (see Ref. \cite{CheMen08} for details). Hence, the modeled pressure Hessian tensor $\textbf{P} = -\frac{\mbox{Tr}(\textbf{A}^2)}{\mbox{Tr}(\textbf{C}_{{\tau_k}}^{-1})}
\textbf{C}_{{\tau_k}}^{-1}$ can be highly anisotropic, mirroring the deformation undergone by 
the fluid as represented by the inverse of $\textbf{C}_{{\tau_k}}$. 
If one is interested in the velocity gradient tensor
of a `coarse-grained' velocity field (such as occurs in large-eddy simulations at length-scale $\Delta$, or in the context of ``tetrads'' when they span a typical scale $\Delta$ in the inertial range), 
the Lagrangian stochastic model may also be interpreted as a model for the coarse-grained velocity gradient tensor if the Kolmogorov time-scale 
$\tau_k$ is replaced by the corresponding eddy turn-over time at the appropriate scale  $\tau_\Delta=\epsilon^{-1/3}\Delta^{2/3}$, where $\epsilon$ is the mean dissipation. In Eq. \ref{eq:Determourmodel}, the viscous term
includes the time-scale $T$, corresponding to the Lagrangian integral time-scale of the velocity  \cite{CheMen06}. 

The model (Eq. \ref{eq:Determourmodel}), consisting of 9 (8 independent) stochastic differential equations, 
can be run with arbitrary initial conditions and it generates stationary statistics for the velocity gradient tensor elements.
In particular, signals corresponding to the time histories of the antisymmetric part of $A_{ij}$ (the vorticity vector and 
its direction $\textbf{e}_{\omega}(t)$) and of the directions of the strain-rate eigenvectors  $\textbf{e}_i(t)$   
are readily obtained from the model runs.  The correlation functions are then evaluated  by averaging signals over long ($10^6 ~T$) records of model signals.

In the following, as comparison we will also make use of data
from a standard direct numerical simulation (DNS) of the
Navier-Stokes equations. DNS is based on a
pseudo-spectral (de-aliased according to the $\frac{3}{2}$-rule) method with 2nd-order accurate Adams-Bashforth time stepping; the computational box is cubic
(size $2\pi$) with periodic boundary conditions in the three
directions and spatial resolution of $256^3$. Statistical stationarity
is maintained by an isotropic external force acting
at low wavenumbers in order to ensure a constant  power
injection. It provides, in the units of the simulation,
a constant energy injection rate of $\epsilon= 0.001$. The kinematic
viscosity of the fluid is $\nu = 0.0004$. The Kolmogorov
scale is $\eta_K =  0.016$ so that $dx/\eta_K \approx 1.5$ (with $dx = 2\pi/256$).  The Taylor-based Reynolds
number is of order $R_\lambda \sim 125$.

The Lagrangian stochastic  model is run with the parameter $\tau_k/T=0.1$, which is appropriate for modeling turbulence with $R_\lambda  \sim  150$  \cite{CheMen08}. 
We compare in Fig. \ref{fig:CorrelAlignVortStr}(a) and (e) the  correlation functions $C_i(\tau)$ as predicted from the model runs (Eq. \ref{eq:Determourmodel}) and measured from DNS. For both the DNS and the model, time is normalized by the ``integral" correlation time $\tau_\eta$ of a single tensor component (e.g. $A_{11}$), namely
$$ \tau_\eta = \frac{1}{\langle A_{11}^2\rangle} \int_0^{+\infty}  \langle A_{11}(t)A_{11}(t+\tau)\rangle d\tau\mbox{ .}$$
It is found that in DNS, $ \tau_\eta\approx 2\tau_k$ with $\tau_k=\sqrt{\nu/\epsilon}$ and for the model, $ \tau_\eta\approx 1.2\tau_k$ with $\tau_k/T=0.1$. We see first that the general trends observed in DNS, and as reported in Ref. \cite{XuPum11} from DNS and for inertial-range tetrads, are reproduced quite well by the stochastic model (Eq. \ref{eq:Determourmodel}). More specifically, for time lags $\tau \lesssim \tau_\eta$ we see that  $\textbf{e}_1$ and $\textbf{e}_\omega$ are increasingly correlated, meaning that we observe, at short times, a more and more pronounced alignment of vorticity with the most extensive eigen-direction of the strain-rate tensor. This was one of the main observations reported in Ref. \cite{XuPum11}. The general trend of initial alignment with the most extensive strain-rate direction  is expected from vortex stretching, as discussed (e.g.)  in Majda (1991) \cite{Maj91}.  Conversely, $C_3(\tau)$ decreases at short times, i.e. vorticity decorrelates with the most contractive direction. As far as the intermediate eigen-direction is concerned, $C_2(\tau)$ decreases monotonically towards the uncorrelated limiting value $C_i(\tau) = 1/3$.  The model shows a slightly faster decay compared to the DNS, in the units of $\tau_{\eta}$.
 
\begin{figure*}[t]
\center{\epsfig{file=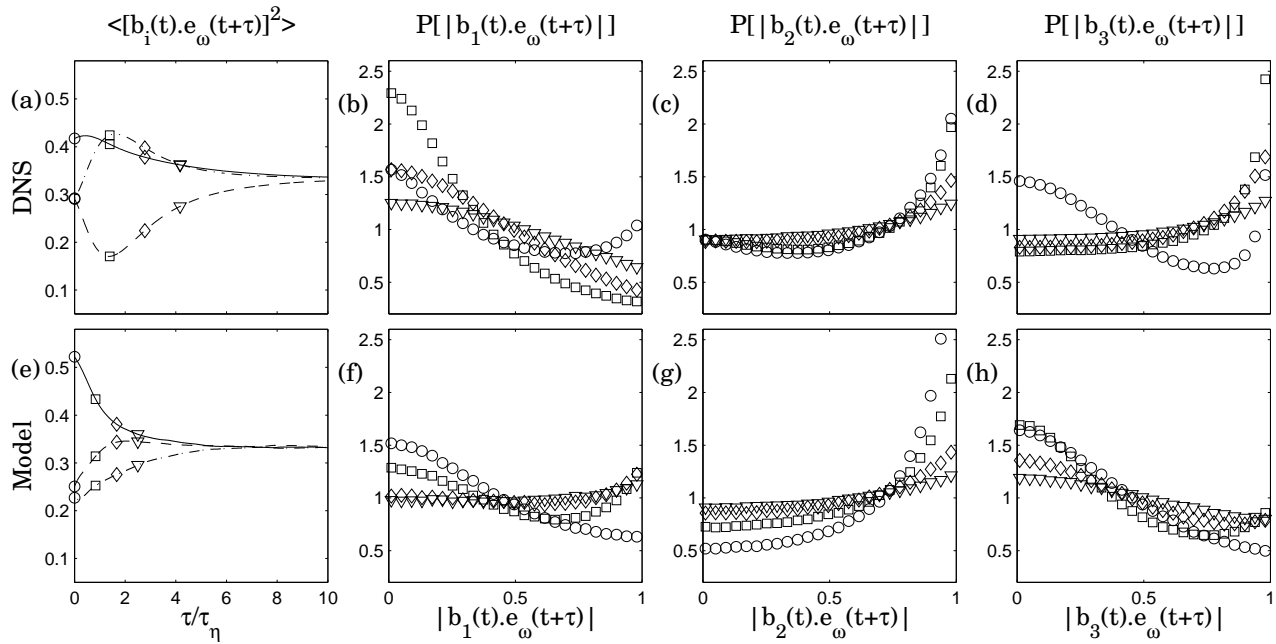,width=17cm}} \caption{(a) and (e) Intercorrelation $C_i(\tau)$ (Eq. \ref{eq:Cit}) of the vorticity direction $\textbf{e}_{\omega}(t)$ and the eigenframe $\textbf{b}_i(t)$ of the deviatoric part of the pressure Hessian $\textbf{P}^d$  in DNS flows and from the model (Eq. \ref{eq:Determourmodel}): dashed line for maximal eigenvalue $\beta_1$, solid line for intermediate eigenvalue $\beta_2$), and dot-dashed line for most negative eigenvalue $\beta_3$.  
Panels (b,c,d) and (f,g,h) are PDFs of $|\textbf{b}_i(t).\textbf{e}_{\omega}(t+\tau)|$ at various time lag $\tau$ represented by different symbols, i.e. $0=\tau_{\circ}<\tau_{\square}<\tau_{\diamond}<\tau_{\triangledown}$ and indicated in (a) and (b). } \label{fig:CorrelAlignVortPH}
\end{figure*}

As proposed in Ref. \cite{XuPum11}, we also display in Figs. \ref{fig:CorrelAlignVortStr}(b,c,d) for DNS, and in Figs.  \ref{fig:CorrelAlignVortStr}(f,g,h) for the model,  the  probability density functions (PDFs) of the cosine of the angles between the direction of vorticity at time $t+\tau$ and the three initial eigen-directions of the deformation at time $t$. Remarkably,  the trends and results observed in DNS are reproduced quite well by the model. At vanishing time-lags $\tau=0$ (symbols $\circ$), $\textbf{e}_\omega$ is uncorrelated with $\textbf{e}_1$, mostly aligned with $\textbf{e}_2$ and mostly orthogonal to $\textbf{e}_3$. Then, for the first non vanishing time lag $\tau_\square$, we observe that $\textbf{e}_\omega(t+\tau_\square)$  becomes more aligned with $\textbf{e}_1(t)$, less aligned with $\textbf{e}_2(t)$ and more orthogonal to $\textbf{e}_3(t)$. For later time lags $\tau_\diamond<\tau_\triangledown$, alignments of vorticity with the eigenframe relax towards the uncorrelated situation, in which the PDF is flat.

We  see thus that the alignments of vorticity with the eigen-frame of the deformation rate tensor are well  predicted by the stochastic Lagrangian model (Eq. \ref{eq:Determourmodel}). A key reason is that the model contains the exact ``velocity gradient self-stretching term'' $-{\textbf{A}}^2$ in its right hand side. This term includes the familiar vortex stretching mechanism that determines the antisymmetric part of the tensor and hence the evolution of vorticity. Therefore, it may be expected that other Lagrangian stochastic models such as the  model with prescribed log-normal dissipation  \cite{Girimajipope90}, the ``tetrad model'' \cite{ChePum99} or the ``Lagrangian linear diffusion model'' \cite{JeoGir03} that include the velocity gradient self-stretching term, should also display the trend of growing alignment of vorticity direction with the most extensive strain eigen-direction. 

Another important and related question is the time correlation of vorticity with the pressure Hessian eigen-frame. The pressure Hessian $\textbf{P}$ is a key quantity that enters into the Lagrangian dynamics of the velocity gradient tensor $\textbf{A}$ (Eq. \ref{eq:TranspA}) \cite{Men11}. As underlined in the literature \cite{CheMen08,Ohk93,OhkKis95,NomPos98,HorFuj08}, understanding the time evolution of the alignments of vorticity with the eigenframe of the deviatoric part of the pressure Hessian $\textbf{P}^d = \textbf{P}-\mbox{tr}(\textbf{P})\textbf{I/3}$, in both the Euler and Navier-Stokes equations, is important. It has been reported that in stationary turbulent flows \cite{CheMen08}, vorticity gets preferentially aligned with the eigenvector associated to the intermediate eigenvalue of $\textbf{P}^d$. This property was found well reproduced by the stochastic model.

We display  in Fig. \ref{fig:CorrelAlignVortPH} the Lagrangian time behavior of alignments of vorticity with the eigen-frame of $\textbf{P}^d$, in a similar way as in 
Fig. \ref{fig:CorrelAlignVortStr} but now with 
the deformation-rate tensor $\textbf{S}$ replaced by $\textbf{P}^d$.  We define the  correlation function
\begin{equation} \label{eq:Dit}
~~D_i(\tau) = \langle \left[ \textbf{b}_i(t).\textbf{e}_{\omega}(t+\tau) \right]^2 \rangle \mbox{ ,}
\end{equation}
where $\textbf{b}_i(t)$ for $i\in\{1,2,3\}$ are the eigenvectors of $\textbf{P}^d$ and the time is, again, understood to be along Lagrangian trajectories. In DNS  (Fig. \ref{fig:CorrelAlignVortPH}(a)), we see that at short time (i.e. $\tau \lesssim \tau_\eta$), $D_1$ (resp. $D_3$) decreases (resp. increases), in the opposite way as it was observed for the deformation-rate tensor (the $C_i$'s). As far as the intermediate pressure Hessian eigendirection is concerned, $D_2$ increases slightly. For larger time lags, i.e.  $\tau \gtrsim \tau_\eta$, all the functions $D_i(\tau)$ relax towards the uncorrelated value $D_i(\tau)=1/3$. In the stochastic model (i.e. in Fig.  \ref{fig:CorrelAlignVortPH}(e)), it seems that even at short times, the functions $D_i(\tau)$ tend immediately from their initial values towards the uncorrelated value of $1/3$: there is no intermediate non-trivial short time behavior. 
 
In Figs.  \ref{fig:CorrelAlignVortPH}(b,c,d) (resp. Figs.  \ref{fig:CorrelAlignVortPH}(f,g,h)), we display the PDFs of the angle of vorticity with the $\textbf{b}_i$ for various time lags. At vanishing time lag $\tau_{\circ}=0$, the PDFs are similar to the ones already displayed in Ref. \cite{CheMen08}. In particular, in both DNS and in the model, the preferential alignment of vorticity and $\textbf{b}_2$ is confirmed. See Ref.  \cite{CheMen08} for further interpretations and discussions about this behavior. As seen in Figs. \ref{fig:CorrelAlignVortPH}(b,d) and  Figs. \ref{fig:CorrelAlignVortPH}(f,h), at short times, vorticity becomes orthogonal to $\textbf{b}_1$ and becomes aligned to  $\textbf{b}_3$. This is not reproduced by the model in which we observe a simple relaxation towards the uncorrelated case (all the PDFs become flat).  Recall that the model is based on assuming that the upstream Lagrangian pressure Hessian is isotropic, and that the tensor ${\bf A}$ does not change during the
(recent) time $\tau_K$. These assumptions are likely to be violated in real turbulence thus causing the observed differences. Note, however, that the evolution of vorticity alignment does not directly depend upon the (symmetric) pressure Hessian tensor, but only indirectly through its dependence on strain-rate. Thus, elucidating the real causes of the observed model limitations is challenging and beyond the scope of this Letter.

To conclude, we have shown in this Letter that at short times, vorticity becomes aligned with the most extensive eigendirection. This agrees with the trends observed for tetrads obtained  at larger scales in  the experimental and DNS investigation in Ref. \cite{XuPum11}. This behavior is quite consistent with linear stretching of the vorticity field. The observed behaviors can be successfully predicted by the  Lagrangian stochastic model based on a closure of the Pressure Hessian using the Recent Fluid Deformation approximation \cite{CheMen06}. We have furthermore extended the study to the alignments of vorticity with the eigen-frame of the pressure Hessian, and we have underlined what is and what is not predicted by the model.

We are very thankful to Emmanuel L\'ev\^eque for providing us with the DNS data. Computations have been performed  by using the local 
computing facilities (PSMN) at ENS Lyon  under grant CPER-CIRA. CM acknowledges support from the National Science Foundation (grant \# CBET-1033942)

\end{document}